\documentclass[prb,aps, twocolumn, amsmath,amssymb,superscriptaddress]{revtex4}
\usepackage{gensymb}
\usepackage{float}
\usepackage{amsmath}
\usepackage{amssymb}
\usepackage{amsfonts}
\usepackage{euscript}
\usepackage{enumerate}
\usepackage{hhline}
\usepackage{pslatex}
\usepackage{tabularx}
\usepackage[usenames,dvipsnames]{xcolor}
\usepackage[colorlinks,linkcolor=red,anchorcolor=blue,citecolor=blue]{hyperref}

\usepackage{graphicx}% Include figure files
\usepackage{dcolumn}% Align table columns on decimal point
\usepackage{bm}% bold math
\usepackage[sort&compress]{natbib}
\usepackage{balance}
\usepackage{float}
\usepackage{amsmath}
\usepackage{amssymb}
\usepackage{amsfonts}
\usepackage{euscript}
\usepackage{enumerate}
\usepackage{hhline}
\usepackage{pslatex}
\usepackage{tabularx}
\usepackage[usenames,dvipsnames]{xcolor}

\usepackage{graphicx}% Include figure files
\usepackage{dcolumn}% Align table columns on decimal point
\usepackage{bm}% bold math

\makeatletter

\renewcommand{\@biblabel}[1]{#1. }
\renewcommand{\@dotsep}{500}
\renewcommand{\@pnumwidth}{0em}
\renewcommand{\l@figure}[2]{% #1 is e.g. Figure 1 + caption, #2 is pg.
\@dottedtocline{1}{1.5em}{2em}{Figure #1}{}\vspace{15pt}}

\begin{document}

\title{Bright Semiconductor Single-Photon Sources Pumped by Heterogeneously Integrated Micropillar lasers with Electrical Injections}

\author{Xueshi Li}
\thanks{These authros contributed equally}
\affiliation{State Key Laboratory of Optoelectronic Materials and Technologies, School of Physics, School of Electronics and Information Technology, Sun Yat-sen University, Guangzhou 510275, China}

\author{Shunfa Liu}
\thanks{These authros contributed equally}
\affiliation{State Key Laboratory of Optoelectronic Materials and Technologies, School of Physics, School of Electronics and Information Technology, Sun Yat-sen University, Guangzhou 510275, China}

\author{Yuming Wei}
\affiliation{State Key Laboratory of Optoelectronic Materials and Technologies, School of Physics, School of Electronics and Information Technology, Sun Yat-sen University, Guangzhou 510275, China}

\author{Jiantao Ma}
\affiliation{State Key Laboratory of Optoelectronic Materials and Technologies, School of Physics, School of Electronics and Information Technology, Sun Yat-sen University, Guangzhou 510275, China}

\author{Changkun Song}
\affiliation{State Key Laboratory of Optoelectronic Materials and Technologies, School of Physics, School of Electronics and Information Technology, Sun Yat-sen University, Guangzhou 510275, China}

\author{Ying Yu}
\affiliation{State Key Laboratory of Optoelectronic Materials and Technologies, School of Physics, School of Electronics and Information Technology, Sun Yat-sen University, Guangzhou 510275, China}

\author{Rongbin Su}
\affiliation{State Key Laboratory of Optoelectronic Materials and Technologies, School of Physics, School of Electronics and Information Technology, Sun Yat-sen University, Guangzhou 510275, China}

\author{Wei Geng}
\affiliation{Hisilicon Research, Huawei Techologies Co., Ltd, Shenzhen 518129, China}

\author{Haiqiao Ni}
\affiliation{State Key Laboratory for Superlattice and Microstructures, State Key Laboratory on Integrated Optoelectronics, Institute of Semiconductors, Chinese Academy of Sciences, Beijing 100083, China}
\affiliation{Center of Materials Science and Optoelectronics Engineering, University of Chinese Academy of Sciences, Beijing 100049, China}

\author{Hanqing Liu}
\affiliation{State Key Laboratory for Superlattice and Microstructures, State Key Laboratory on Integrated Optoelectronics, Institute of Semiconductors, Chinese Academy of Sciences, Beijing 100083, China}
\affiliation{Center of Materials Science and Optoelectronics Engineering, University of Chinese Academy of Sciences, Beijing 100049, China}

\author{Xiangbin Su}
\affiliation{State Key Laboratory for Superlattice and Microstructures, State Key Laboratory on Integrated Optoelectronics, Institute of Semiconductors, Chinese Academy of Sciences, Beijing 100083, China}
\affiliation{Center of Materials Science and Optoelectronics Engineering, University of Chinese Academy of Sciences, Beijing 100049, China}

\author{Zhichuan Niu}
\thanks{zcniu@semi.ac.cn}
\affiliation{State Key Laboratory for Superlattice and Microstructures, State Key Laboratory on Integrated Optoelectronics, Institute of Semiconductors, Chinese Academy of Sciences, Beijing 100083, China}
\affiliation{Center of Materials Science and Optoelectronics Engineering, University of Chinese Academy of Sciences, Beijing 100049, China}

\author{Youling Chen}
\thanks{ylchen@semi.ac.cn}
\affiliation{State Key Laboratory for Superlattice and Microstructures, State Key Laboratory on Integrated Optoelectronics, Institute of Semiconductors, Chinese Academy of Sciences, Beijing 100083, China}
\affiliation{Center of Materials Science and Optoelectronics Engineering, University of Chinese Academy of Sciences, Beijing 100049, China}

\author{Jin Liu}
\thanks{liujin23@mail.sysu.edu.cn}
\affiliation{State Key Laboratory of Optoelectronic Materials and Technologies, School of Physics, School of Electronics and Information Technology, Sun Yat-sen University, Guangzhou 510275, China}

\date{\today}% It is always \today, today,

\begin{abstract}
\noindent \textbf{
The emerging hybrid integrated quantum photonics combines advantages of different functional components into a single chip to meet the stringent requirements for quantum information processing. Despite the tremendous progress in hybrid integrations of III-V quantum emitters with silicon-based photonic circuits and superconducting single-photon detectors, on-chip optical excitations of quantum emitters via miniaturized lasers towards single-photon sources (SPSs) with low power consumptions, small device footprints and excellent coherence properties is highly desirable yet illusive. In this work, we present realizations of bright semiconductor singe-photon sources heterogeneously integrated with on-chip electrically-injected microlasers. Different from previous one-by-one transfer printing technique implemented in hybrid quantum dot (QD) photonic devices, multiple deterministically coupled QD-circular Bragg Grating (CBG) SPSs were integrated with electrically-injected micropillar lasers at one time via a potentially scalable transfer printing process assisted by the wide-field photoluminescence (PL) imaging technique. Optically pumped by electrically-injected microlasers, pure single photons are generated with a high-brightness of a count rate of 3.8 M/s and an extraction efficiency of 25.44\%. Such a high-brightness is due to the enhancement by the cavity mode of the CBG, which is confirmed by a Purcell factor of 2.5. Our work provides a powerful tool for advancing hybrid integrated quantum photonics in general and boosts the developments for realizing highly-compact, energy-efficient and coherent SPSs in particular.}
\end{abstract}

\maketitle
\section{Introduction}
Photonic quantum technology harnessing the superposition and entanglement of non-classic states of light has enabled secure communication, superfast computation and accurate metrology~\cite{o2009photonic}. Introductions of integrated optics to modern quantum photonics facilitate the abilities of generations, manipulations and detections of quantum states of light by using more than 1,000 programable components on a phase-stable chip with a millimetre-scale footprint~\cite{wang2020integrated}. Similar to conventional integrated optics, it is not realistic for a single material platform to satisfy all the functionalities required for each distinct component in quantum information processing tasks~\cite{elshaari2020hybrid,kim2020hybrid}. For example, III-V semiconductor QD SPSs so far have exhibited the best performances in terms of simultaneous high-degrees of source brightness, single-photon purity and photon indistinguishability~\cite{senellart2017high,wang2019towards,tomm2021bright,he2017deterministic,uppu2020scalable} while silicon photonic circuits excel in reconfigurable manipulations of multi-photon~\cite{qiang2018large,li2020metalens} or high-dimensional quantum states~\cite{wang2018multidimensional,zhang2019generation} with high operational fidelities. Thus, hybrid integrated quantum photonic circuits are currently being under extensive explorations worldwidely by heterogeneously integrating III-V solid-state quantum emitters as deterministic quantum light sources with silicon-based photonic circuits~\cite{davanco2017heterogeneous,zadeh2016deterministic,katsumi2019quantum,katsumi2018transfer} and superconducting single-photon detectors~\cite{schwartz2018fully,najafi2015chip,reithmaier2013chip} on a single chip. Despite remarkable progress in the development of the heterogeneously integrated photonic components, there remains a large performance gap between the fully-fledged integrated quantum photonic devices and their classical photonic counterparts—that is—chip-integrated pumping lasers. The on-chip optical excitation of solid-state quantum emitters by using microlasers results in much lower power consumptions, significantly reduced device footprints and potentially ideal coherence properties of the emitted photons. In a conventional off-chip optical excitation scheme, radiations from bulky and high-power lasers have to be greatly attenuated to a few tens of nano-Watt for creating carriers in QDs, in which most of the radiation power of the bulky external lasers is wasted. In addition, optical pumping schemes, especially resonant~\cite{he2013demand}, quasi-resonant~\cite{gschrey2015highly}, two-photon resonant~\cite{wei2022tailoring,jayakumar2013deterministic,reindl2017phonon} and phonon-assisted excitation~\cite{thomas2021bright}, are able to deliver highly-coherent single photons with a near-unity photon indistinguishability, appreciably surpassing the record value of 41\% achieved under the electrical injection condition~\cite{schlehahn2016electrically}.

Moving towards integrations of QD SPSs and microlasers, electrically-injected micropillars have been successfully demonstrated to excite QD SPSs in an off-chip manner~\cite{kreinberg2018quantum}. The on-chip excitations of QDs have also been demonstrated in a monolithic chip with unsatisfactory source brightnesses. In monolithic platforms, both SPSs and microlasers are limited to the same form, e.g., micropillars ~\cite{munnelly2017electrically,stock2013chip} or planar cavities~\cite{lee2017electrically}, which prevents the independent optimizations of each component. To better separate optimizations of laser excitations and single-photon emissions, we combine, in this work, deterministically-fabricated planar CBGs as bright SPSs and electrically-injected micropillars as highly-directional microlasers on a single chip. These two components are individually optimized and heterogeneously integrated together by using a potentially scalable transfer printing process capable of fabricating a multitude of devices in a single run. The single QD was pumped by an on-chip micropillar laser under electrical injections, exhibiting high-performances in terms of the source brightness and single-photon purity thanks to the coupling of the QD to the cavity mode of the CBG. This work constitutes a major step in developing highly-efficient and coherent semiconductor SPSs with small footprints for hybrid integrated quantum photonics.

\begin{center}
	\begin{figure}
		\begin{center}
			\includegraphics[width=\linewidth]{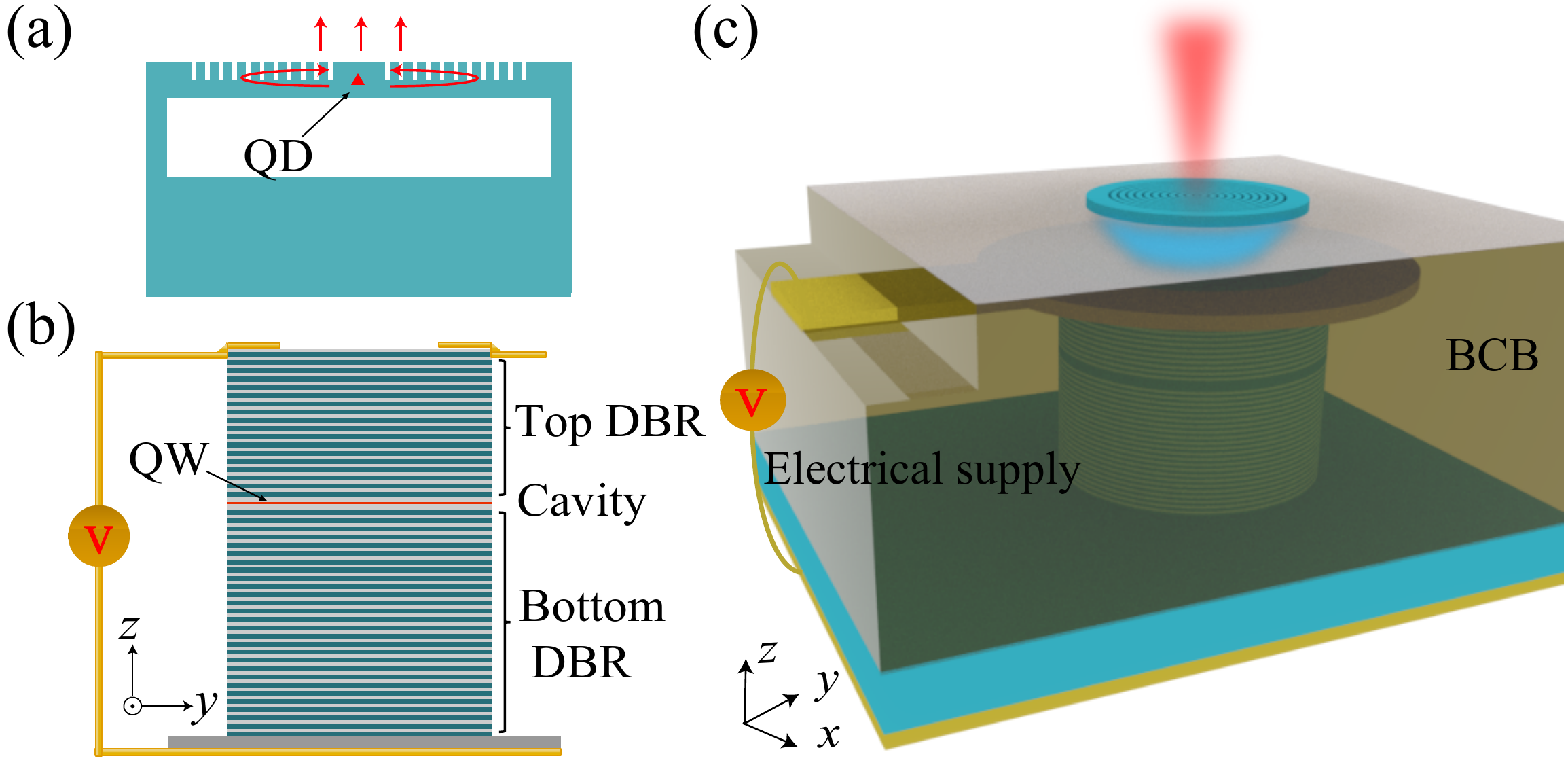}
			\caption{\textbf{Illustration of the hybrid integrated SPS.} (a) CBG with a single QD embedded for bright single-photon emissions. The single photons emitted from the QD are confined in the suspended membrane by total internal reflections, partially reflected to the centre of the CBG and partially scattered to the top by second-order Bragg gratings consisting of shallowly etched air trenches. (b) Micropillars with embedded quantum wells for electrically-injected microlasers. (c) Illustration of the device. Pumped by the on-chip micropillar laser with electrical injections, the QD in the CBG emits highly-directional single photons to the free space.}
			\label{fig:Fig1}
		\end{center}
	\end{figure}
\end{center}

\section{Results}

Figure~1 presents the building blocks and the concept of our device. A suspended GaAs CBG with a single QD embedded in the center serves as an efficient SPS~, as shown in Fig.~1(a). The CBG consists of a GaAs microdisk surrounded by a serial of shallowly etched air trenches. Photons emitted from the QD in the center of CBG are mostly confined in the suspended membrane. Due to the presence of second-order Bragg gratings, parts of the emitted photons from the QD are scattered upwards for efficient collections and the others are reflected back to form a cavity for enhancing the strength of light-matter interactions~\cite{davanco2011circular,sapienza2015nanoscale,wang2019demand,abudayyeh2021single,abudayyeh2021overcoming,nikolay2018accurate}. To optically pump the QD-CBG SPS, we explore electrically-injected micropillar lasers~\cite{gies2019quantum} consisting of quantum wells sandwiched between two distributed Bragg Reflectors (DBRs) [Fig.~1(b)]. These two elements are assembled in a way that the CBG is located right on top of the micropillar laser and an additional spacer is designed specifically to separate cavity modes in each distinct device, as schematically shown in Fig.~1(c). Optically driven by the coherent photons emitted from the electrically-injected micropillar laser, the QD in the CBG emits highly-directional single photons upwards to the collection optics.

With the above device design, we further explore the optical performance of a single-QD in the hybrid cavities by modeling the extraction efficiency and the Purcell factor, as shown in Fig.~2(a). The extraction efficiency is defined as the fraction of the photons collected by the objective over all the photons emitted by the QD. The cavity effect of the CBG is revealed by a Lorentzian shape of the Purcell factor with a maximal value of $\sim$15 and a full width at half maximum of $\sim$3~nm, corresponding to a Q-factor of $\sim$303. The bandwidth for efficient collections of the single photons are much broader, showing extraction efficiencies \textgreater40\% over 33~nm. Such an appreciable Purcell factor is due to the tight light confinement of the CBG cavity mode, as presented in Fig.~2(b). We chose a 2~$\mu$m gap filled with benzocyclobutene (BCB) between the CBG and the micropillar laser so that the cavity mode of the CBG is not perturbed by the presence of the micropillar. The excitation of the QD by the micropillar laser (blue) and the highly-directional single-photon emissions (red) can be clearly identified from the beam propagation profiles in the XZ plane of the device, as shown in the Fig.~2(c). Fig.~2(d) further shows the far-field pattern of single-photon emissions from the CBG device, exhibiting a divergent angle within 10~degree for the efficient collection.

\begin{figure}[!h]
	\includegraphics[width=\linewidth]{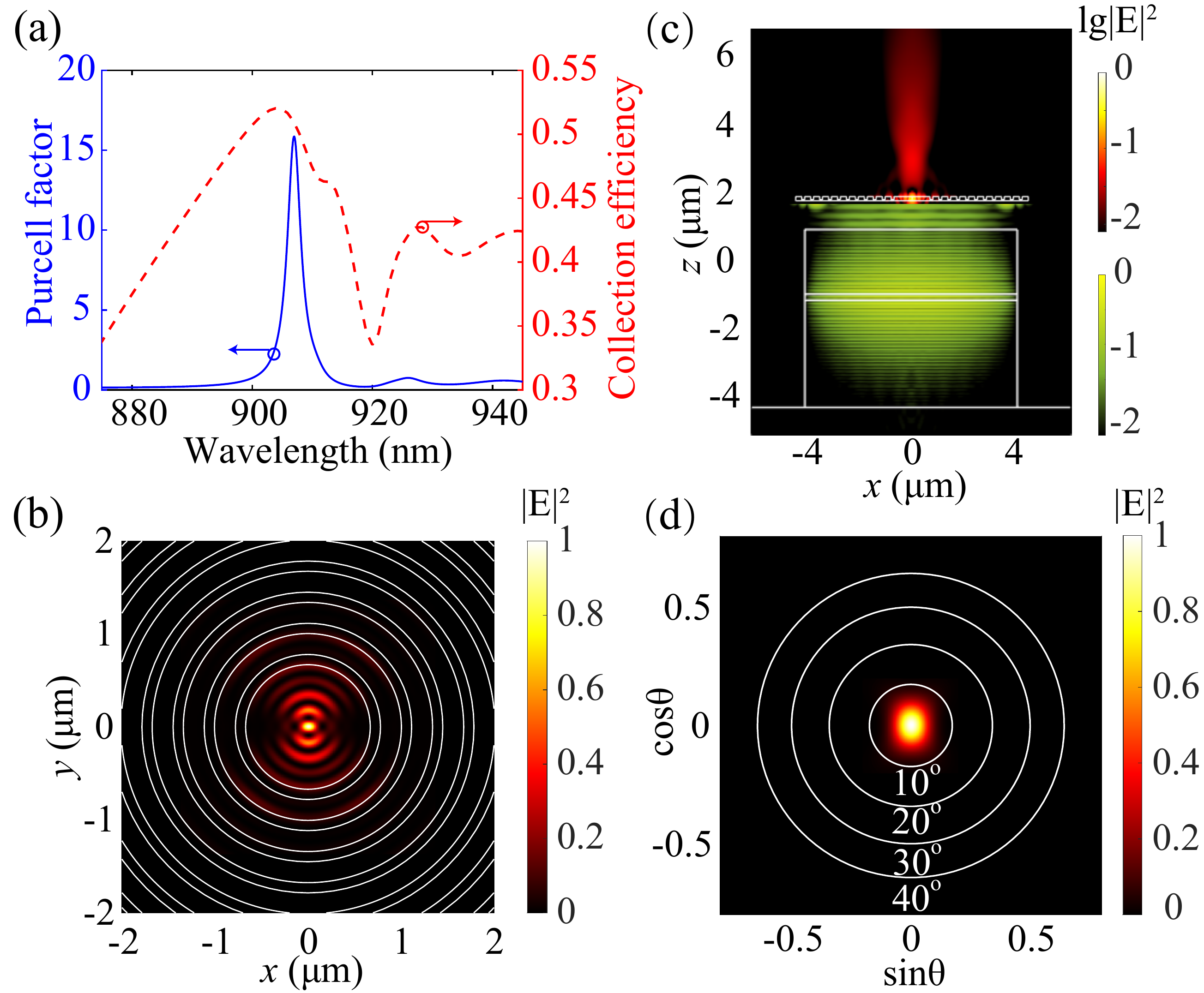}
	\caption{\textbf{Numerical simulations of the device.} (a) Purcell factor and extraction efficiency of the hybrid integrated SPS. A numerical aperture (NA) of 0.65 was used in the extraction efficiency calculation. (b) The near-field intensity distribution of the cavity mode of the CBG shows highly-localized electromagnetic fields. (c) The bean propagations of laser emissions from the micropillar and the single-photon emissions from the CBG in the XZ plane. The QD is efficiently excited by the micropillar laser and emits highly-directional single photons. (d) The far-field intensity distribution of single-photon emissions exhibits a divergent angle within 10 degrees.}
	\label{fig:Fig2}
\end{figure}

\begin{figure*}
	\begin{center}
		\includegraphics[width=0.7\linewidth]{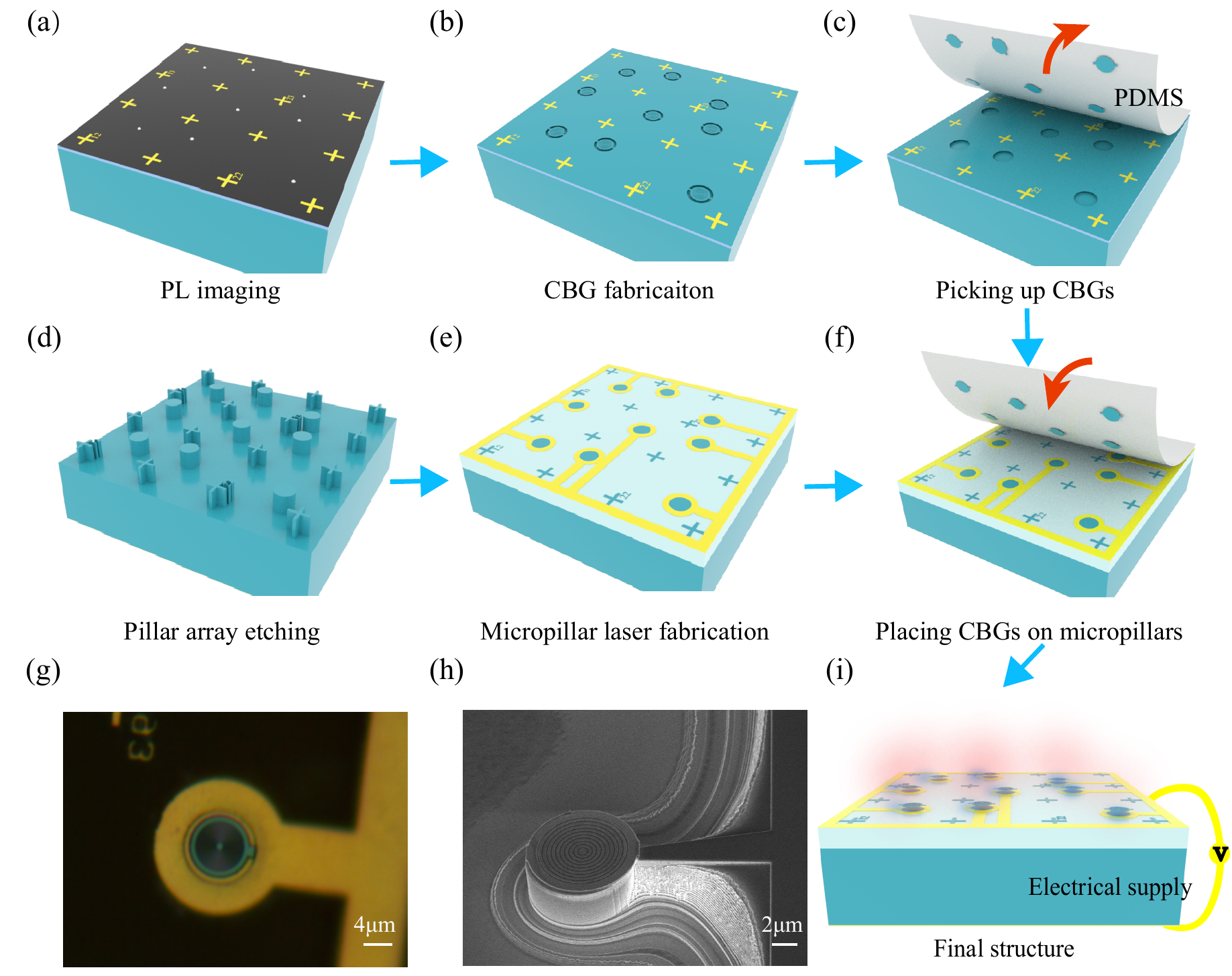}
		\caption{\textbf{The device fabrication flow.} (a) Preparation of alignment marks and the extractions of QD positions via a PL imaging process. (b) Deterministically coupled QD-CBG devices fabricated by an aligned E-beam lithography, a dry etch and a wet etch process. (c) The CBGs are picked up by attaching and quickly peeling off a transparent rubber stamp. (d) Fabrication of a micropillar array based on the relative positions of QDs in (a). (e) Fabrication of micropillar laser arrays with electrodes. (f) Accurately placements of multiple CBGs on top of the micropillar array, resulting in (i) the final structure. The successful transfer with a high accuracy is supported by the microscope image (g) and SEM image (h).}
		\label{fig:Fig3}
	\end{center}
\end{figure*}

We take advantage of the well-developed transfer printing technique which has been successfully demonstrated for building hybrid quantum photonic devices~\cite{katsumi2018transfer,katsumi2019quantum,jin2022generation}. However, the implementations of such a technique to QD devices so far are limited in a one-by-one fashion due to the intrinsic random nature in QD's spatial position. In order to obtain multiple devices with high yields at one time, we utilized the wide-field PL imaging technique to deterministically fabricate coupled CBG-QD devices~\cite{sapienza2015nanoscale,liu2017cryogenic,liu2021nanoscale}. In such a process, metallic alignment marks were firstly prepared on an III-V QD wafer and then the spatial positions of individual QDs respective to the alignment marks were extracted from PL images [Fig.~1(a)]. The deterministically coupled QD SPSs were then obtained by an aligned E-beam lithography, a chloride-based dry etch and a selective wet etch, as presented in Fig.~3(b). Then, we used transfer printing to pick up multiple processed CBGs from the substrate using a transparent rubber stamp [Fig.~3(c)]. On the other III-V wafer containing quantum wells and DBRs, electrically-injected microlasers were fabricated in a specific array based on the relative positions of fabricated CBGs, with the similar processing recipe for micropillar SPSs~\cite{liu2021dual,wei2022tailoring}, as shown in Fig.~3(d,e). The micropillar array was covered with a $ \sim $2$~\mu$m thick BCB spacer between the microlaser and SPS in order to maintain the cavity modes of the on-substrate CBGs. Because the cavity mode slightly leak out of the thin membrane in z direction, as shown in Fig.~S(5), direct placements of CBGs on tops of micropillars result in the vanish of CBG modes, significantly reducing the brightness of the SPSs. We then aligned multiple CBGs to the micropillar laser array under an optical microscope and released the CBGs on micropillars by slowly and gently peeling the stamp off [Fig~3(f)]. A schematic of the completed hybrid integrated SPS array is shown in Fig.~3(i), in which multiple QDs coupled to CBGs can be simultaneously pumped by the underneath electrically-injected micropillar laser array. In a representative microscope image of the transferred device [Fig.~3(g)], a CBG can be clearly identified in the center of the micropillar laser. The successful transfer printing with a high alignment accuracy was further supported by a scanning electron microscope (SEM) image of a test device without planarizing the micropillar laser using BCB, as presented in Fig.~3(h).

The optical setup and the device characterizations are shown in Fig.~4. The hybrid integrated devices are placed on three-dimensional nanopositioners located in a closed circle cryostat with a base temperature of 5.4~K. The QD SPSs can be either excited by the on-chip micropillar lasers or by external continuous wave (CW) and pulsed lasers via a 50 X objective with a numerical aperture of 0.65. The emitted single photons are collected by the same objective and sent to a monochromator for the spectral analysis or to single-photon detectors for lifetime characterizations and Hanbury-Brown-Twiss (HBT) measurements, as shown in Fig.~4(a). By applying a small bias voltage to the micropillar laser, single sharp excitonic lines (red) from the single QD are identified in the $\mu$PL spectrum, as presented in Fig.~4(b). The bright exciton (X1) is spectrally close to the cavity mode (black) of the CBG. The emission spectrum of the electrically-injected micropillar laser is shown in the inset of Fig.~4(b), featuring a sharp lasing peak at 854 nm. Such near-infrared micropillar lasers are compatible with industry-standard, which renders practical application of our hybrid QDs SPS-Micropillar laser device convenient. The coupling between the QD and the CBG was revealed by the shortened lifetime of X1. A lifetime of 529~ps for QD in the CBG was extracted, corresponding to a Purcell factor of 2.5 compared to a lifetime of 1316 ps for the QD in bulk. The large deviation of the simulated Purcell factor from the experimental result is a combination of the non-ideal QD position and the long carrier relaxation time under above-band excitation~\cite{liu2018high}.

\begin{center}
	\begin{figure}
		\includegraphics[width=\linewidth]{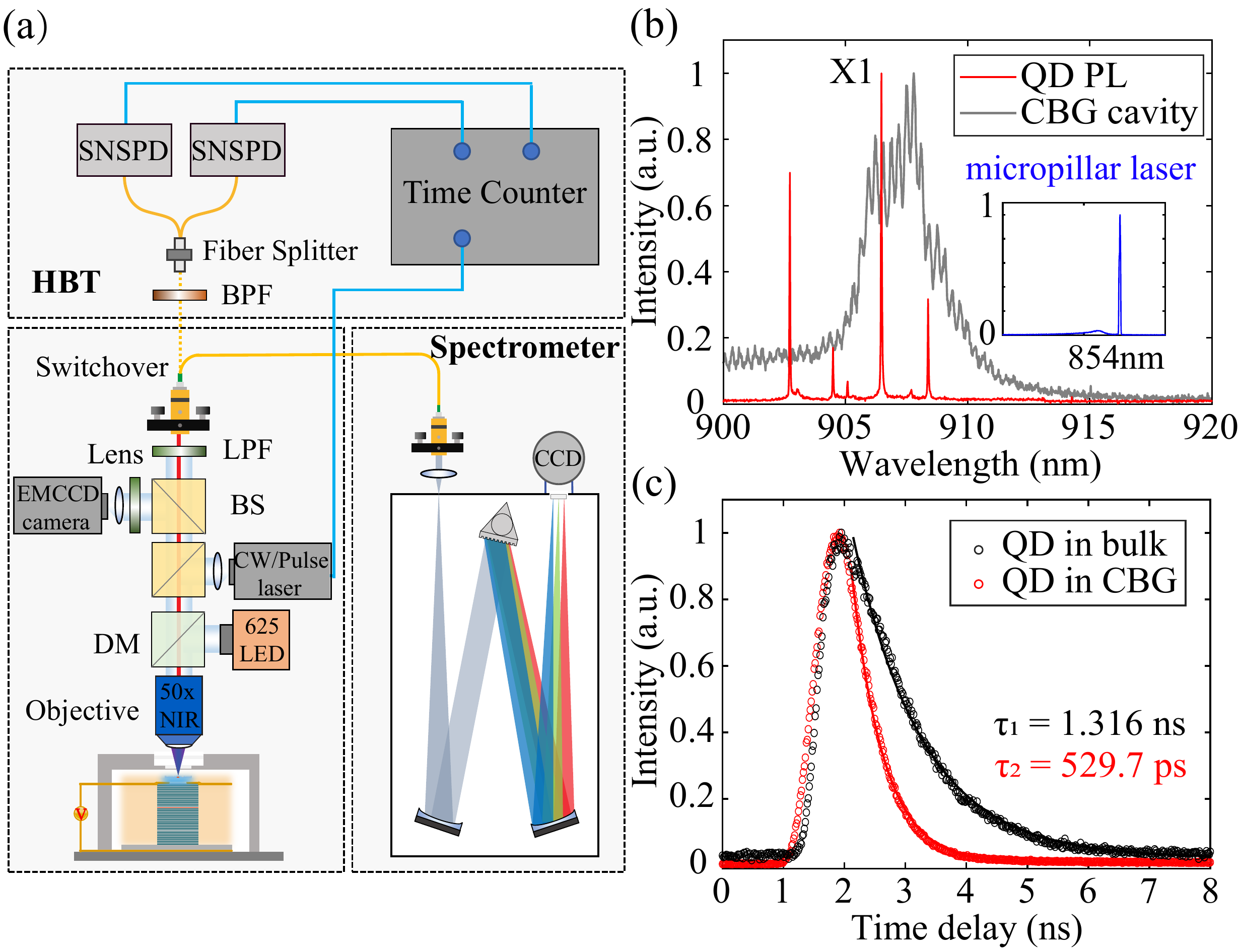}
		\caption{\textbf{Experiment setup and device characterizations.} (a) Schematic of the experiment setup. BS: beam spliter. LPF: long-pass filter. DM: dichroic mirror. BPF: band-pass filter. SNSPD: superconducting nanowire single-photon detector. (b) PL spectrum of excitonic states associated with a single QD (red line) and the cavity mode of CBG (black line). (c) Comparison of the lifetimes for the QD coupled to the CBG (red) and a QD in bulk (black).}
		\label{fig:Fig4}
	\end{figure}
\end{center}

\begin{center}
	\begin{figure}[h]
		\includegraphics[width=\linewidth]{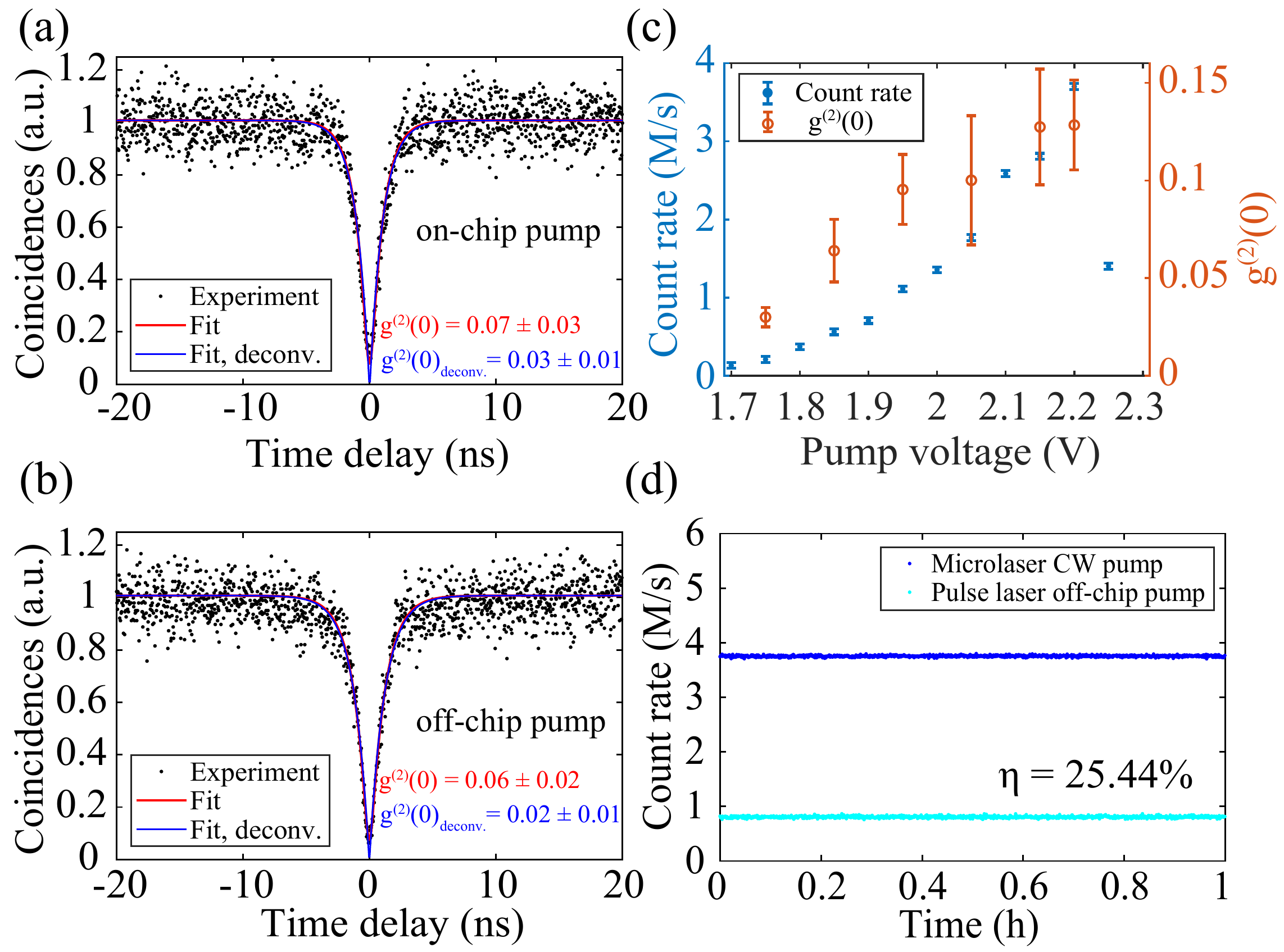}
		\caption{\textbf{Source brightness and single-photon purity.}  (a, b) Second-order correlation as a function of time delay for the QD pumped by the on-chip microlaser and an external CW laser diode. Black dots mark experimental data. The red and blue curves are the fits with and without deconvolutions of the detector response, respectively. (c) Photon count rate (blue circles) and g$^{2}$(0) (brown circles) as a function of the bias voltage applied to the micropillar laser. (d) Saturation count rates from X1 pumped by the micropillar laser (dark blue) operating in a CW mode and a fs pulsed laser (bright blue).}
		\label{fig:Fig5}
	\end{figure}
\end{center}

A single-photon purity g$^2$(0) as low as 0.03$\pm$0.01 (deconvoluted with the detector response~\cite{davanco2017heterogeneous}) under on-chip excitation condition was extracted from the second-order correlation measurement in an HBT interferometer, see Fig.~5(a). As a comparison, we used an external CW diode laser at 854~nm to excite the QD and obtained a very similar g$^2$(0) of 0.02$\pm$0.01 [Fig.~5(b)], which suggests that the on-chip optical excitation by integrated lasers can deliver the same performance as that obtained under the widely-used external optical pumping. The photon count rate and g$^2$(0) as a function of the bias voltage are shown in Fig.~5(c). Both photon flux and g$^2$(0) increase~\cite{madsen2014efficient,kolatschek2021bright, xu2022bright} with the rise of the applied bias voltage and the maximal count rate (not saturated) up to 3.8~M/s with a g$^2$(0) of 0.13$\pm$0.02 was obtained under the voltage of 2.2~V. Further increase of the bias voltage results in a rapid quenching of single-photon emissions, which is probably due to the detrimental thermal effect associated with the BCB spacer. We further quantified the extraction efficiency of the hybrid integrated SPS by externally exciting the device using a femto-second (fs) pulsed laser with a repetition rate of 79.6~MHz. The saturated count rate under pulsed excitation (see Fig.~S10) together with the count rate under on-chip CW pumping are presented in Fig.~5(d). The generated photon rate is monitored  continuously over 1 hour to check the long-term operational stability of the device. Under the pulsed excitation, each laser pulse is presumably to generate one single photon~\cite{lodahl2015interfacing}. By measuring the count rate in the single-photon detector and carefully calibrating the system detection efficiency (see details in SI), an extraction efficiency of 25.44\% was obtained for the hybrid integrated SPSs.

\section{Discussion}

To conclude, we have demonstrated hybrid integrations of bright SPSs with on-chip electrically-injected micropillar lasers. Assisted by the QD PL imaging technique, our optimized transfer printing process operates at a multi-device level, showing potential scalability for hybrid integrated quantum photonics. The single QD in the CBG was optically pumped by an on-chip micropillar laser, exhibiting a CW photon count rate of 3.8 M/s with g$^2$(0) of 0.13$\pm$0.02 under a high bias voltage and a very low g$^2$(0) of 0.03$\pm$0.01 under a low bias voltage. Our device experienced a Purcell factor of 2.5 and an extraction efficiency of 25.44\% thanks to the coupling of QD to the CBG cavity mode. This work serves as a crucial step towards energy-saving on-chip single-photon and entangled pair sources for hybrid integrated quantum photonics. Our devices could be simultaneously coupled to an optical fiber array to realize the plug-and-play function, which has broad application prospects in quantum key distribution. Moving forwards, on-chip resonant excitations (see a realistic proposal in SI) by spectrally tuning the microlaser wavelength to the QD transition energies could be pursued to further improve the photon indistinguishability for advancing photonic quantum technologies.

\section{Methods}

\noindent\textbf{Epitaxial growth of QDs: }

The samples were grown using solid source molecular beam epitaxy on semi-insulating GaAs (001) substrates. After deoxidization at a temperature of 680 $ ^\circ C $ for 10 minutes and growth of 300 nm GaAs buffer layer at 660 $ ^\circ C $, a 1000 nm $\rm{Al_{0.8}Ga_{0.2}As}$ sacrificial layer was grown at 620 $ ^\circ C $. The InAs QDs were embedded in the middle of a GaAs layer with a thickness of 160 nm. The InAs QDs were deposited at the temperature of (Tc-27) $ ^\circ C $ with an indium flux rate of 0.004 ML/s and an As flux pressure of 5x$\rm{10^{-7}}$ Torr.  The deposition temperatures are calibrated by the transition temperature Tc when the surface reconstruction pattern of GaAs in high-energy electron diffraction (RHEED) transfers from (2 × 4) to (2 × 3). Then the InAs QDs were capped with a thin layer of 0.3 nm AlAs and 6.5 nm GaAs, followed by a 200s indium flushing step at 660 X.

\noindent\textbf{Fabrication of the coupled QD-CBG: }

The process starts with the III-V wafer consists of a 160 nm-thin GaAs membrane containing InAs quantum dots (QDs) grown on top of a sacrificial layer (1~$\rm{\mu}$m $\rm{Al_{0.8}Ga_{0.2}As}$) and a GaAs substrate. To acquire the positions of QDs, firstly, metallic alignment marks are created on the surface of the sample with standard E-beam lithography, metal deposition and lift-off processes. Then, the positions of the QDs respective to the alignment marks are extracted from the wide-field PL images. The PL imaging process has an accuracy of about 20~nm~\cite{he2017deterministic}. After that, the shallow etched CBGs with QD in the center are fabricated through another aligned E-beam lithography followed by a chlorine-based dry etch process. After removing the photoresist by oxygen plasma surface treatments, we dip the sample in the acid solution of HF 10\% for the necessary time to remove the sacrificial layer. We finally obtain the suspended deterministically coupled QD-CBG devices by drying the sample in isopropanol. The detailed fabrication flow of the coupled QD-CBG devices is presented at Supplementary Fig. S1.

\noindent\textbf{Fabrication of the micropillar laser: }

The wafer consists of a single layer GaAs with QW between 23(30) top(bottom) GaAs/$\rm{Al_{0.9}Ga_{0.1}As}$ distributed Bragg reflector (DBRs) provided by EPIHOUSE. In order to achieve an electrically pumped micropillar laser, we need to properly dop the wafers. The epitaxial n-type and p-type regions are realized by doping the GaAs during the growth with silicon and carbon, respectively. The first step in creating the micropillar laser is the fabrication of electrical contacts to the n-doped layers. We obtain the Ni/Ge/Au/Ni/Au contacts on the back of the wafer by E-beam evaporation, followed by an annealing process. Then, the masks of the electrically-injected micropillar lasers and alignment marks are fabricated in a specific array based on the relative positions of fabricated CBGs by using an E-beam lithography. After the chlorine-based dry etch process, we use BCB to flatten the micropillars to ensure that the upper surfaces of the micropillars are just exposed. This process requires multiple spin-coatings of the photoresist, annealing and dry etching processes. Finally, the positive electrode is created by using an E-beam lithography, metal evaporation and lift-off processes. The detailed fabrication flow of the micropillar laser is presented at Supplementary Fig.~S2.

\noindent\textbf{Transfer printing process: }

After the fabrications of the CBGs, we realize the micropillars at the corresponding positions according to the layout of the CBG array. Therefore, during the transfer process, we can align multiple CBGs with micropillars to achieve potentially scalable device integrations.We show three of 8 transferred devices at one time in Supplementary Fig. S3.

\vspace{16pt}
\noindent \textbf{Acknowledgements: }This research was supported by National Key Research and Development Program of China (2018YFA0306103); National Natural Science Foundation of China (11874437, 62035017); Hisilicon Technologies CO., LIMITED and the national super-computer center in Guangzhou.

\noindent\textbf{Conflit of interest: }The authors declare no conflict of interest.

\noindent\textbf{Author contributions: }Y.~L.~C and J.~L conceived the project. X.~S.~L, R.~B.~S and S.~F.~L performed numerical simulations. C.~K.~S, Y.~Y, H.~Q.~N, H.~Q.~L, X.~B.~S and Z.~C.~N grew the wafers. X.~S.~L, J.~T.~M and S.~F.~L performed optical characterizations. X.~S.~L, W.~G, S.~F.~L and J.~L analyzed the data. J.~L wrote the manuscript with inputs from all authors. Z.~C.~N, Y.~L.~C and J.~L supervised the project.

%\bibliographystyle{naturemag}
%\bibliography{bibfile}% Produces the bibliography via BibTeX.

%\newpage
\clearpage
\onecolumngrid \bigskip

\begin{center} {{\bf \large SUPPLEMENTARY INFORMATION}}\end{center}

\setcounter{figure}{0}
\setcounter{section}{0}
\makeatletter
\renewcommand{\thefigure}{S\@arabic\c@figure}

\section{Characterization of the QD wafer}
The QD wafer is schematically shown in Fig.~S1(a), consisting of a 160 nm-thin GaAs membrane containing InAs quantum dots (QDs), a sacrificial layer (1~$\rm{\mu}$m $\rm{Al_{0.8}Ga_{0.2}As}$) and a GaAs substrate. Fig.~S1(b) shows the high-resolution high-angle annular dark field (HAADF) STEM images along [110] crystallographic direction of single InAs QD. The bright regions close to the center of the image represent the InAs QDs, with the average height of ~6 nm. The AlGaAs capping layer can be clearly presented as the darker region surrounding the QD. A representative spectrum of single QD is shown in Fig.~S1(c).
\begin{figure}[!h]
	%\begin{center}
	%\includegraphics[width=\linewidth]{SIFigure1.pdf}
	\includegraphics[width=0.7\linewidth]{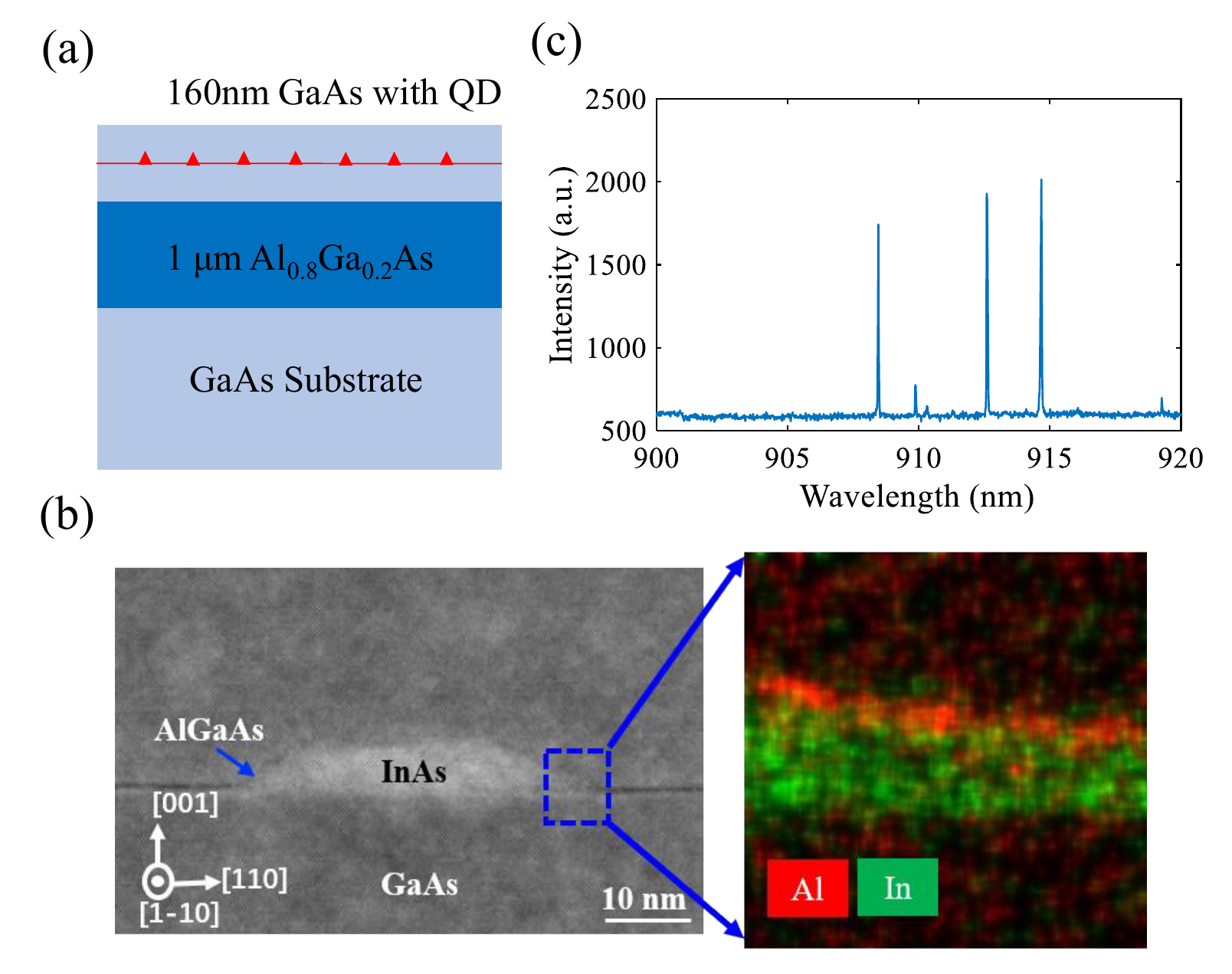}
	\caption{Characterization of the QD wafer. (a) Schematics of the QD wafer. (b) STM image of the InAs QDs used in the work. (c). A representative PL spectrum for the single QD.}
	\label{SIfig:Fig1}
	%\end{center}
\end{figure}

\newpage
\section{Fabrication flow of the CBG}

The full fabrication flow of deterministically coupled QD-CBG is shown in Fig.~S2(a). The process starts with the III-V wafer consists of a 160 nm-thin GaAs membrane containing InAs quantum dots (QDs) grown on top of a sacrificial layer (1$\rm{\mu}$m $\rm{Al_{0.8}Ga_{0.2}As}$) and a GaAs substrate. To acquire the positions of QDs, firstly, metallic alignment marks are created on the surface of the sample with standard E-beam lithography, metal deposition and lift-off processes. Then, the positions of the QDs respective to the alignment marks are extracted from the wide-field PL images. After that, the shallow etched CBGs with QD in the center are fabricated through another aligned E-beam lithography followed by a chlorine-based dry etch process. After removing the photoresist by oxygen plasma surface treatments, we dip the sample in the acid solution of HF 10\% for the necessary time to remove the sacrificial layer. We finally obtain the suspended deterministically coupled QD-CBG devices by drying the sample in isopropanol, as shown in Fig.~S2(b). The parameters of CBG during fabrication are carefully chosen to ensure the spectral matching between the X state of the QD and the fundamental cavity mode of the CBG.

\begin{figure}[!h]
	%\begin{center}
	%\includegraphics[width=\linewidth]{SIFigure1.pdf}
	\includegraphics[width=0.7\linewidth]{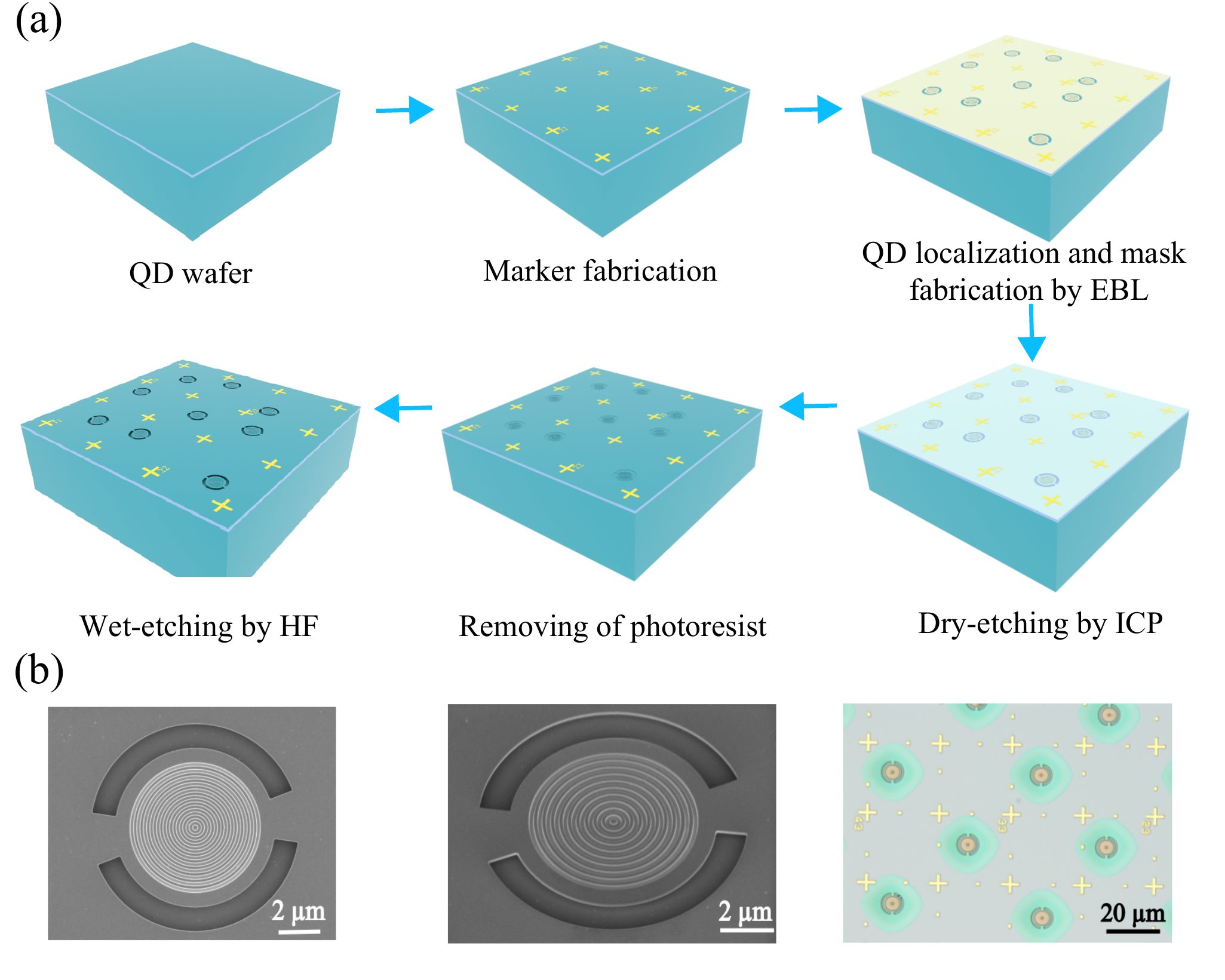}
	\caption{Fabrication flow of deterministically coupled QD-CBG devices.}
	\label{SIfig:Fig1}
	%\end{center}
\end{figure}

\newpage
\section{Fabrication flow of the micropillar laser}

The full fabrication flow of the micropillar laser is shown in Fig.~3(a). The wafer consists of a single layer GaAs with QW between 23(30) top(bottom) GaAs/$\rm{Al_{0.9}Ga_{0.1}As}$ distributed Bragg reflector (DBRs) provided by EPIHOUSE. In order to achieve an electrically pumped micropillar laser, we need to properly dop the wafers. The epitaxial n-type and p-type regions are realized by doping the GaAs during the growth with silicon and carbon, respectively. The first step in creating the micropillar laser is the fabrication of electrical contacts to the n-doped layers. We obtain the Ni/Ge/Au/Ni/Au contacts on the back of the wafer by E-beam evaporation, followed by an annealing process. Then, the masks of the electrically-injected micropillar lasers and alignment marks are fabricated in a specific array based on the relative positions of fabricated CBGs by using an E-beam lithography. After the chlorine-based dry etch process, we use BCB to flatten the micropillars to ensure that the upper surfaces of the micropillars are just exposed. This process requires multiple spin-coatings of the photoresist, annealing and dry etching processes. Finally, the positive electrode is created by using an E-beam lithography, metal evaporation and lift-off processes. The SEM images of the samples are shown in Fig.~S3(b).

\begin{figure}[!h]
	%\begin{center}
	%\includegraphics[width=\linewidth]{SIFigure1.pdf}
	\includegraphics[width=0.7\linewidth]{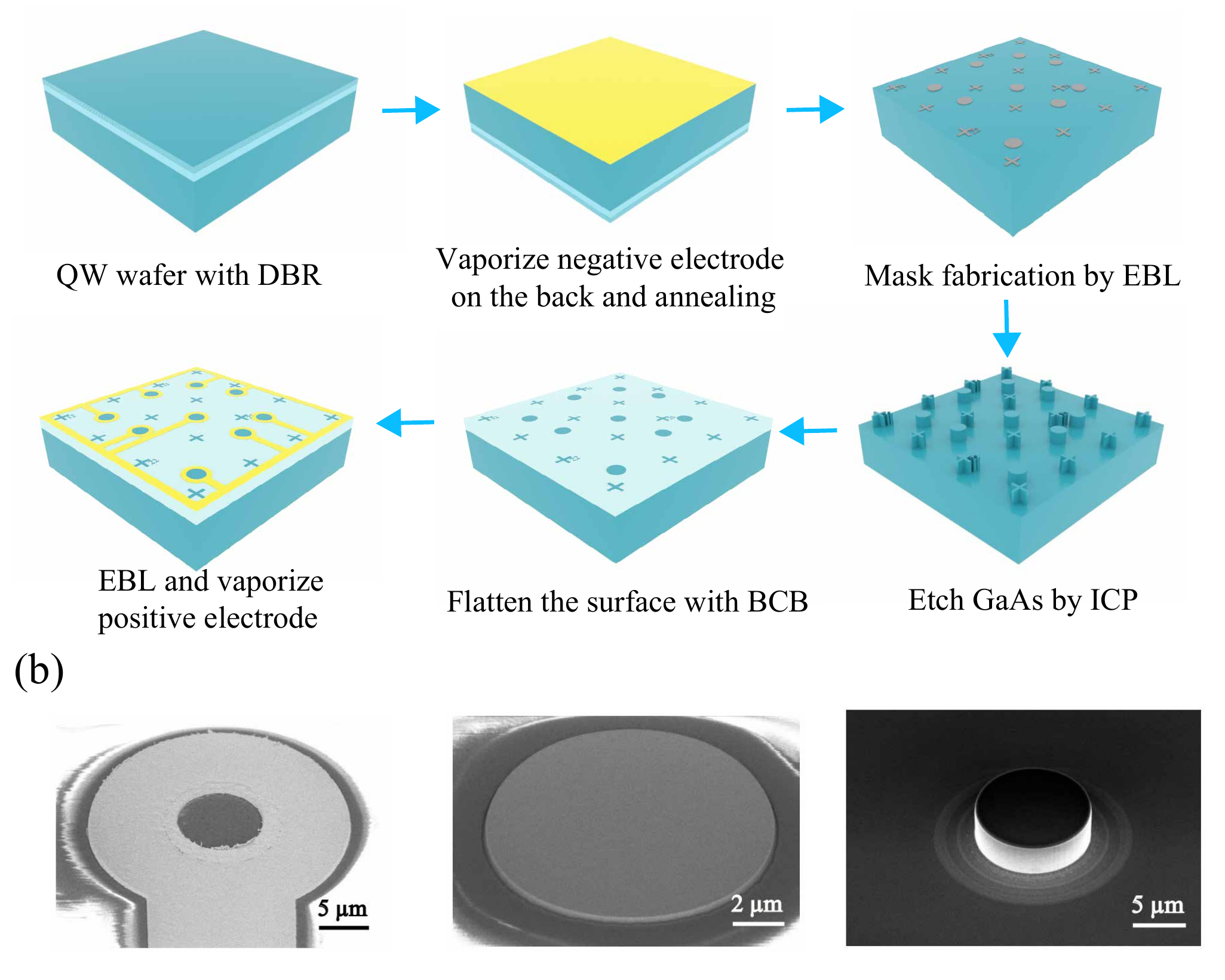}
	\caption{Fabrication flow of micropillar laser.}
	\label{SIfig:Fig3}
	%\end{center}
\end{figure}

\newpage
\section{Transfer printing process for multiple devices}

After the fabrications of the CBGs, we realize the micropillars at the corresponding positions according to the layout of the CBG array. Therefore, during the transfer process, we can align multiple CBGs with micropillars to achieve potentially scalable device integrations. We show three of 8 transferred devices at one time in Fig.~S4(a,b). The extraction efficiencies of each device are listed in Fig.~S4(c), featuring high yields for the multiple device transfer printing process.

\begin{figure}[!h]
	%\begin{center}
	\includegraphics[width=0.7\linewidth]{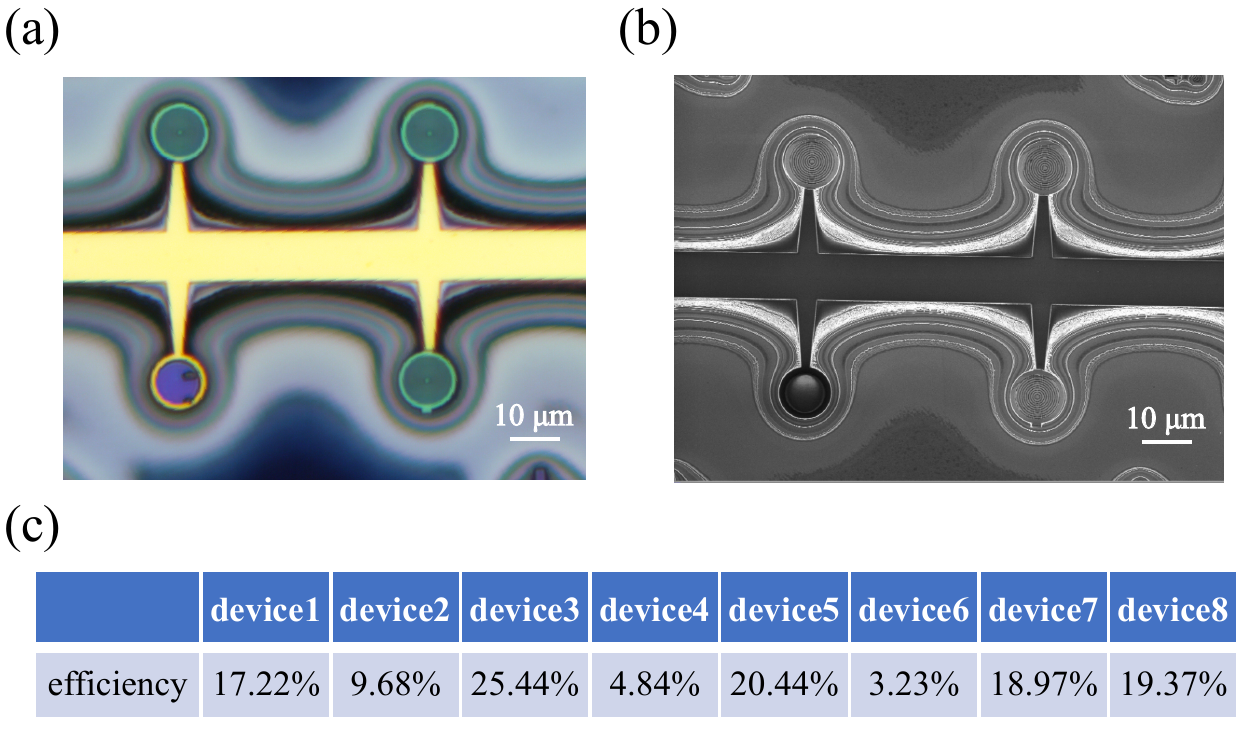}
	\caption{Transfer printing process for multiple devices.}
	\label{SIfig:Fig4}
	%\end{center}
\end{figure}

We further present the transferring of a large array of devices in Fig.~S5. 20 out of 26 devices are successfully fabricated in a single run. The fabrication throughput and yield could be further boosted by exploiting commercially available tools that are specially designed for the transfer printing process.
\begin{figure}[!h]
	%\begin{center}
	\includegraphics[width=0.7\linewidth]{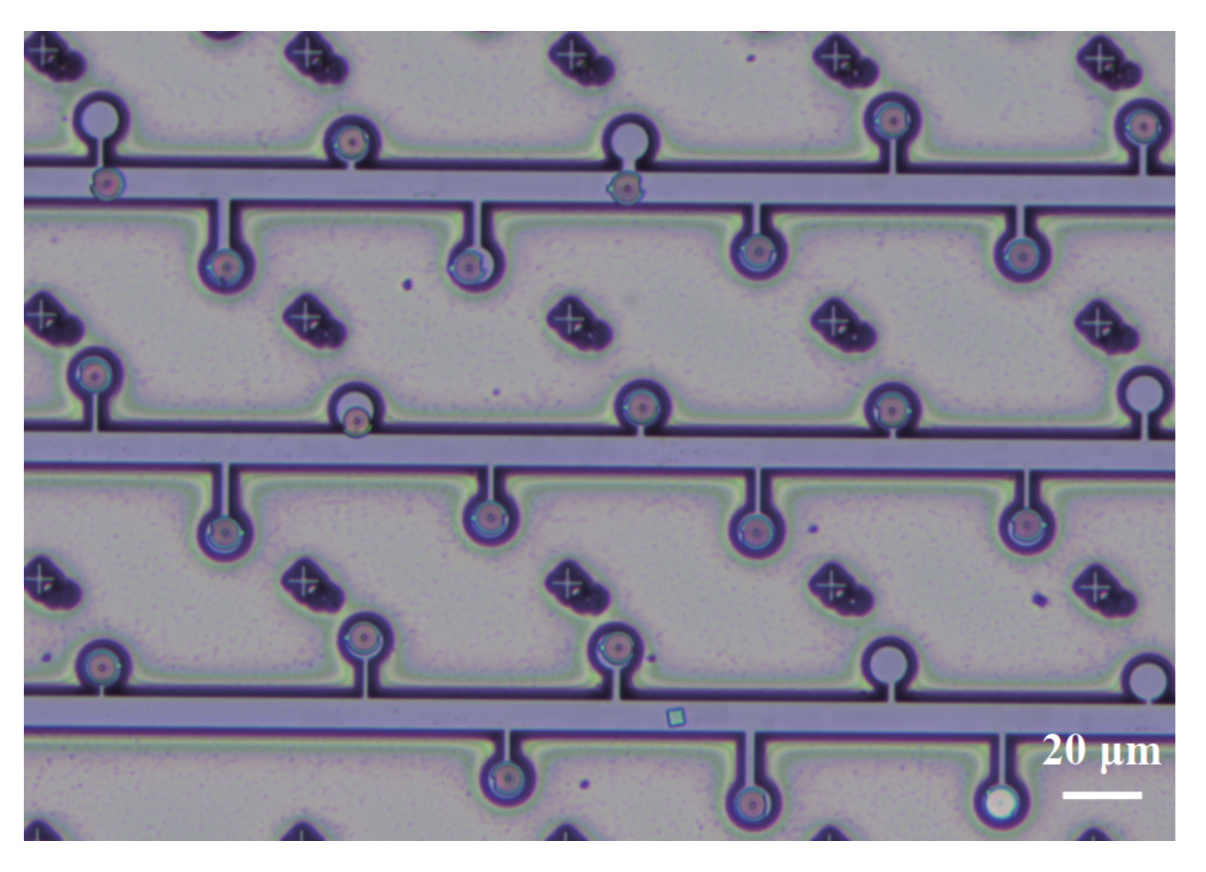}
	\caption{The result of a large scale transfer printing process.}
	\label{SIfig:Fig5}
	%\end{center}
\end{figure}

\newpage
\section{Numerical simulations of CBG}
The design of CBG for QD single-photon sources has been thoroughly discussed in the literature ~\cite{davanco2011circular,sapienza2015nanoscale}. Here we briefly show the optimization of the parameters in Fig.~S(6). We use the central disk with a radius of 2p in which p and w are the periodicity and the width of the etched trenches respectively, as shown in Fig.~S6(a). The cavity modes in XY plane and YZ plane are plotted in Fig.~S6(b,c) respectively. In Fig.~S6(d), the cavity resonances for different p are shown. To match the QD emission wavelength (~907 nm), we chose p=335 nm. We further change w in Fig.~S6(e), and w=110 nm gives rise to the highest Purcell factor at the targeted wavelength. Finally, the extraction efficiency for the optimized parameter (p=335 nm and w=110 nm) as a function of NA is presented in Fig.~S6(f).

We further present the Purcell factor and extraction efficiency (with NA=0.65) as a function of the placement deviation and fabrication resolution in Fig.~S7 and Fig.~S8, respectively. The displacement refers the distance from QD to the center of the CBG. The fabrication resolution mostly reflects on the width of the etched trenches. Due to the uncertainty of the dipole orientation, we have simulated both x-polarized and y-polarized dipoles. The Purcell factors and extraction efficiencies of other orientated dipoles fall into the range between the x-polarized and y-polarized dipoles.~\cite{sapienza2015nanoscale} Compared to the simulations, our QD seems to be within 300 nm from the center of the CBG, which is probably due to an unintentional offset in the aligned electron beam lithography process.

\begin{figure}[!h]
	%\begin{center}
	\includegraphics[width=0.90\linewidth]{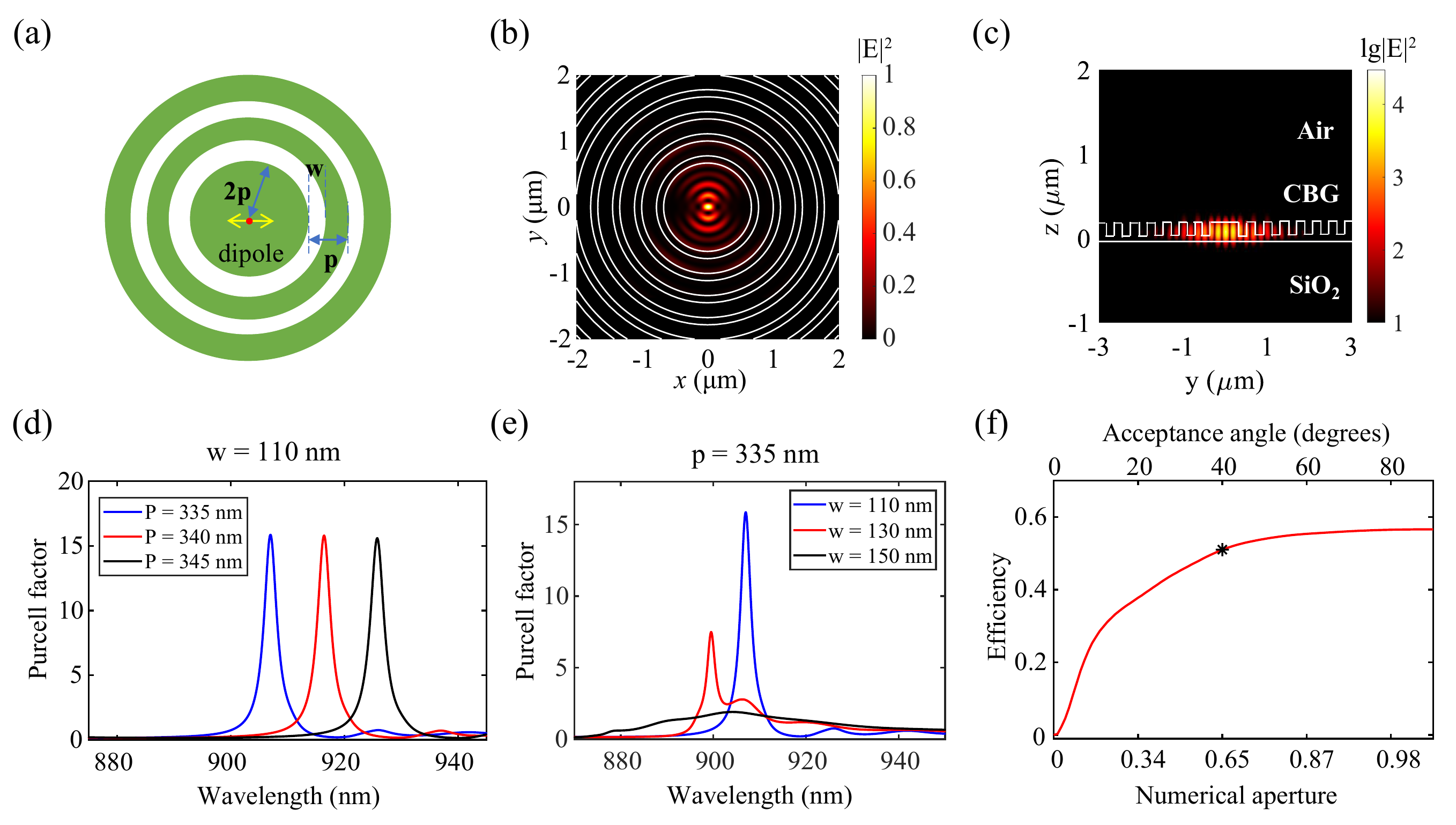}
	\caption{Simulations of the CBG. (a) Schematics of the CBG. (b) Cavity mode profile in XY plane. (c) Cavity profile in YZ plane. (d) Spectra of cavity modes with different p. (e) Spectra of cavity modes with different w. (f) Extraction efficiency as a function of the NA for the CBG with p=335 nm and W= 110 nm. The star denotes the NA used in our experiment.}
	\label{SIfig:Fig6}
	%\end{center}
\end{figure}

\begin{figure}[!h]
	%\begin{center}
	\includegraphics[width=0.85\linewidth]{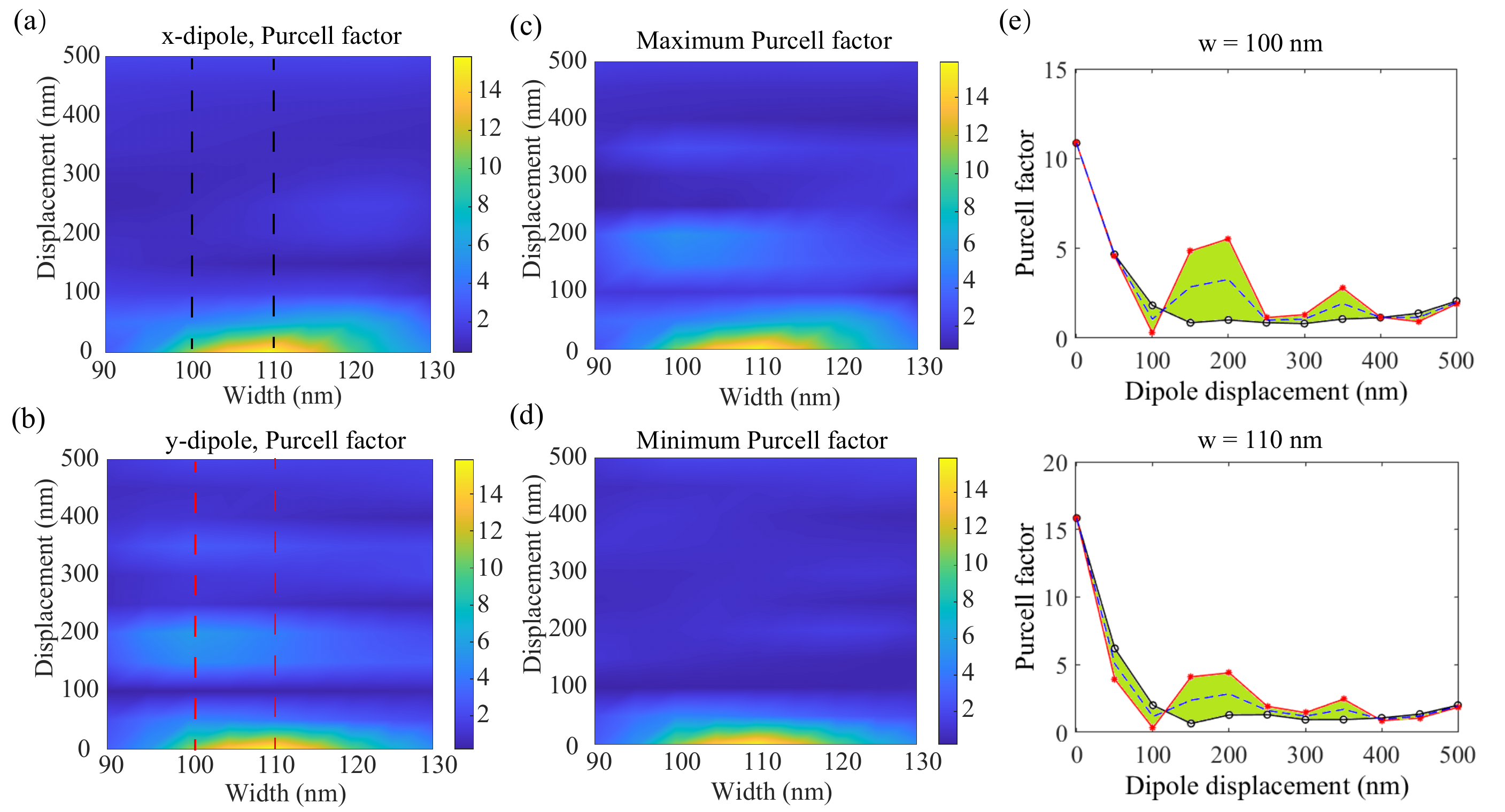}
	\caption{Purcell factor as a function of etched trench width and placement deviation inside the CBG. (a) Results for an x-oriented dipole; (b) Result for a y-oriented dipole; (c) Maximum achievable Purcell factor; (d) Minimum achievable Purcell factor; (e) Purcell factor as a function of dipole displacement from the CBG with w=100 nm and w=110 nm [shown as dashed lines in (a)-(b)]. Shaded areas correspond to the uncertainty in Purcell factor due to lack of knowledge of the dipole orientation. The dashed line corresponds to an average value.}
	\label{SIfig:Fig6}
	%\end{center}
\end{figure}

\begin{figure}[!h]
	%\begin{center}
	\includegraphics[width=0.85\linewidth]{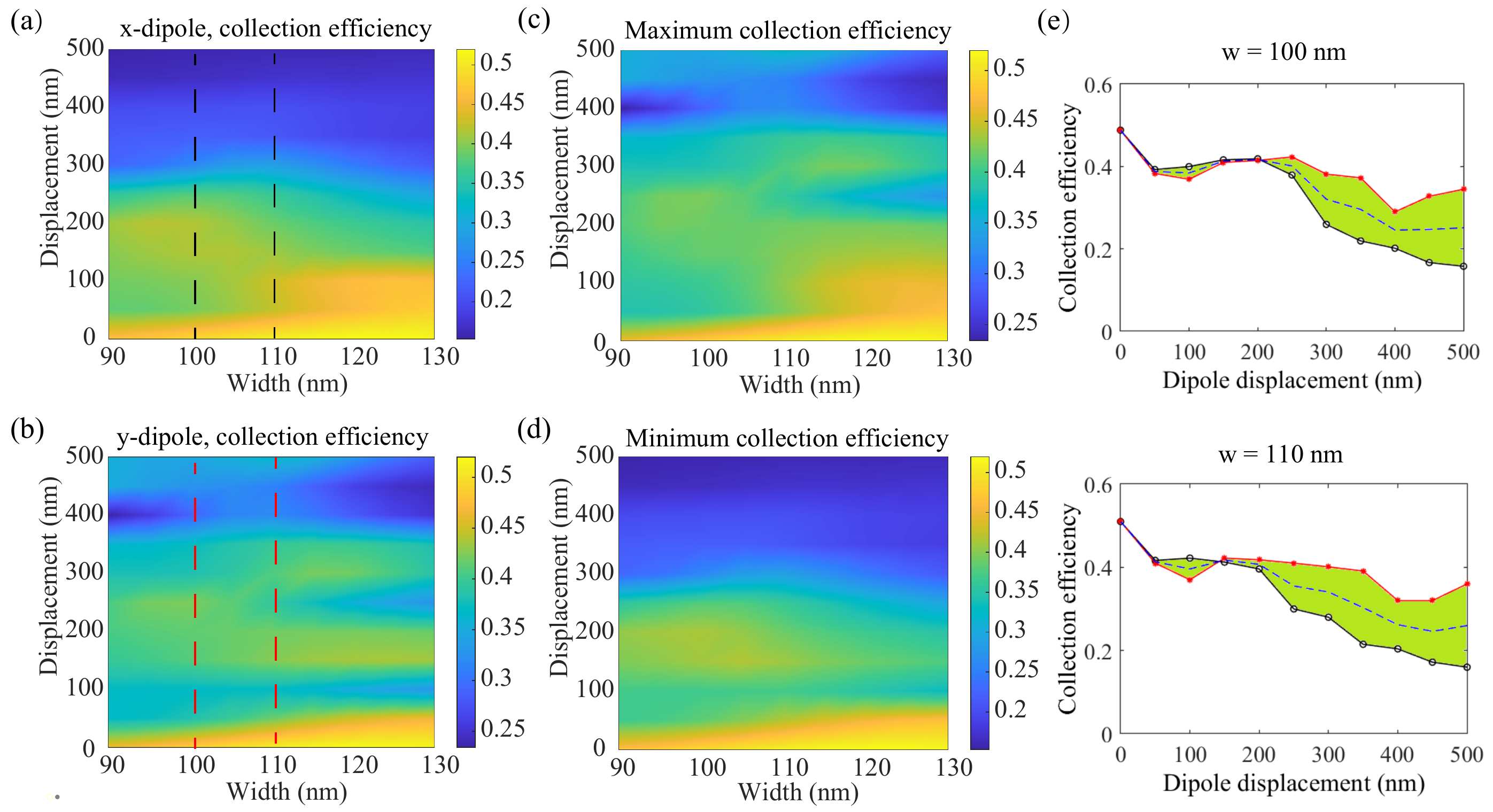}
	\caption{Extraction efficiency (NA=0.65) as a function of etched trench width and placement deviation inside the CBG. (a) Results for an x-oriented dipole; (b) Results for a y-oriented dipole; (c) Maximum achievable extraction efficiency; (d) Minimum achievable extraction efficiency; (e) Extraction efficiency as a function of dipole displacement from the CBG with w=100 nm and w=110 nm [shown as dashed lines in (a)-(b)]. Shaded areas correspond to the uncertainty in extraction efficiency due to lack of knowledge of the dipole orientation. The dashed line corresponds to an average value.}
	\label{SIfig:Fig7}
	%\end{center}
\end{figure}

\newpage
\section{Characterizations of microlasers}
We’ve presented the far-field pattern of the microlaser in Fig.~S9. The far-field emission is close to an Gaussian profile with a divergence angle below 10 degree.
\begin{figure}[!h]
	%\begin{center}
	\includegraphics[width=0.9\linewidth]{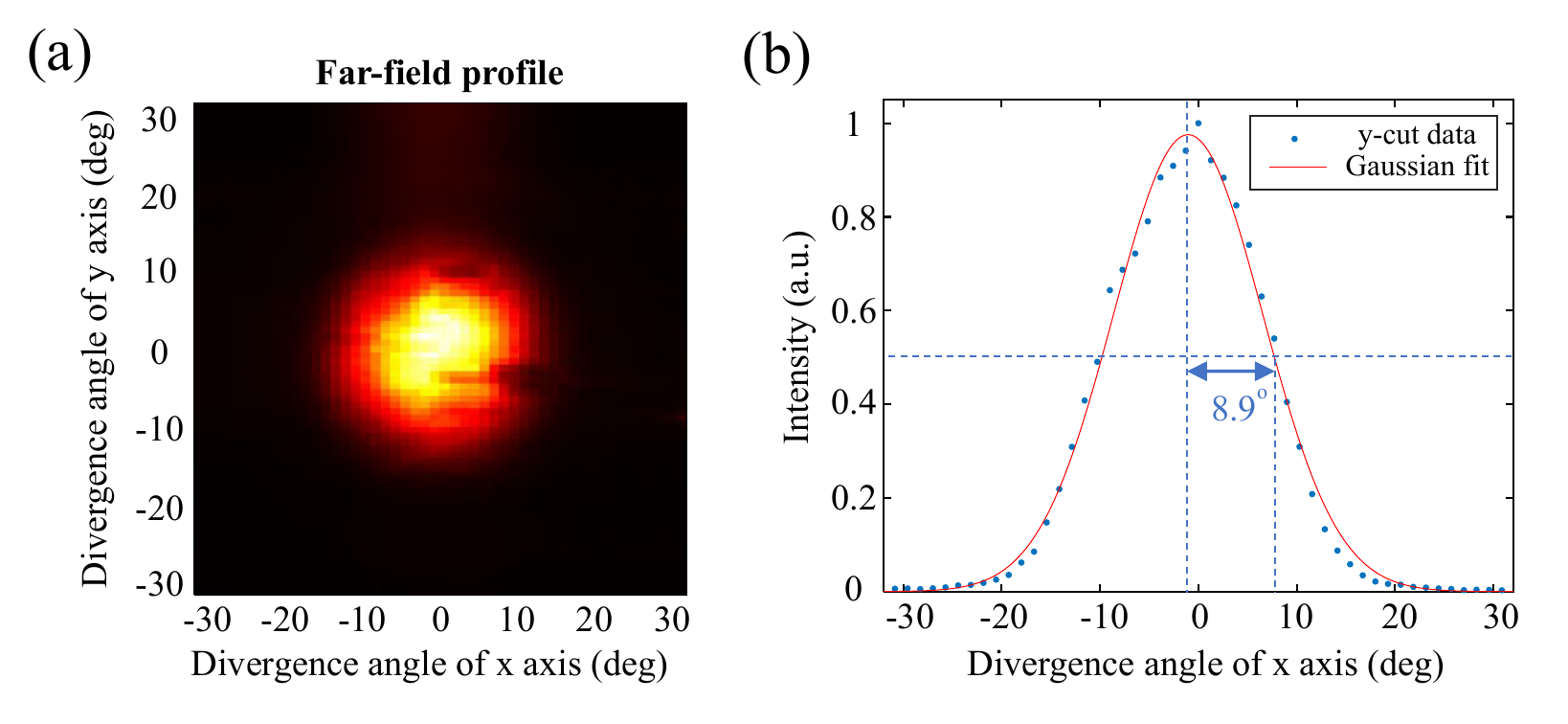}
	\caption{Beam profile of the microlaser. (a) Far-field pattern of the electrically injected micropillar laser. (b) y-cut of (a), exhibiting a divergence angle below 10 degree.}
	\label{SIfig:Fig8}
	%\end{center}
\end{figure}

\newpage
\section{Estimation of extraction efficiency}
The extraction efficiency of the hybrid SPS at the first lens is evaluated by externally exciting the device using a femto-second pulsed laser with a repetition rate of 79.6~MHz. Fig.~S10(a) shows the schematic of the experimental setup for estimating the extraction efficiency. The emitted photons are collected using a microscope objective with an NA of 0.65. For this measurement, the signals pass through a dichroscope (DM), a beam splitter (BS), a long pass filter (LPF), and a bandpass filter(BPF) before coupling into a single-mode fiber. The detection efficiency of the whole setup has been carefully calibrated to be 0.0405, as shown in Fig.~S10(b). A saturation count rate of 0.82 MHz was extracted from Fig.~S10(c), which results in an extraction efficiency at the first lens of 25.44\%.

\begin{figure}[!h]
	%\begin{center}
	\includegraphics[width=\linewidth]{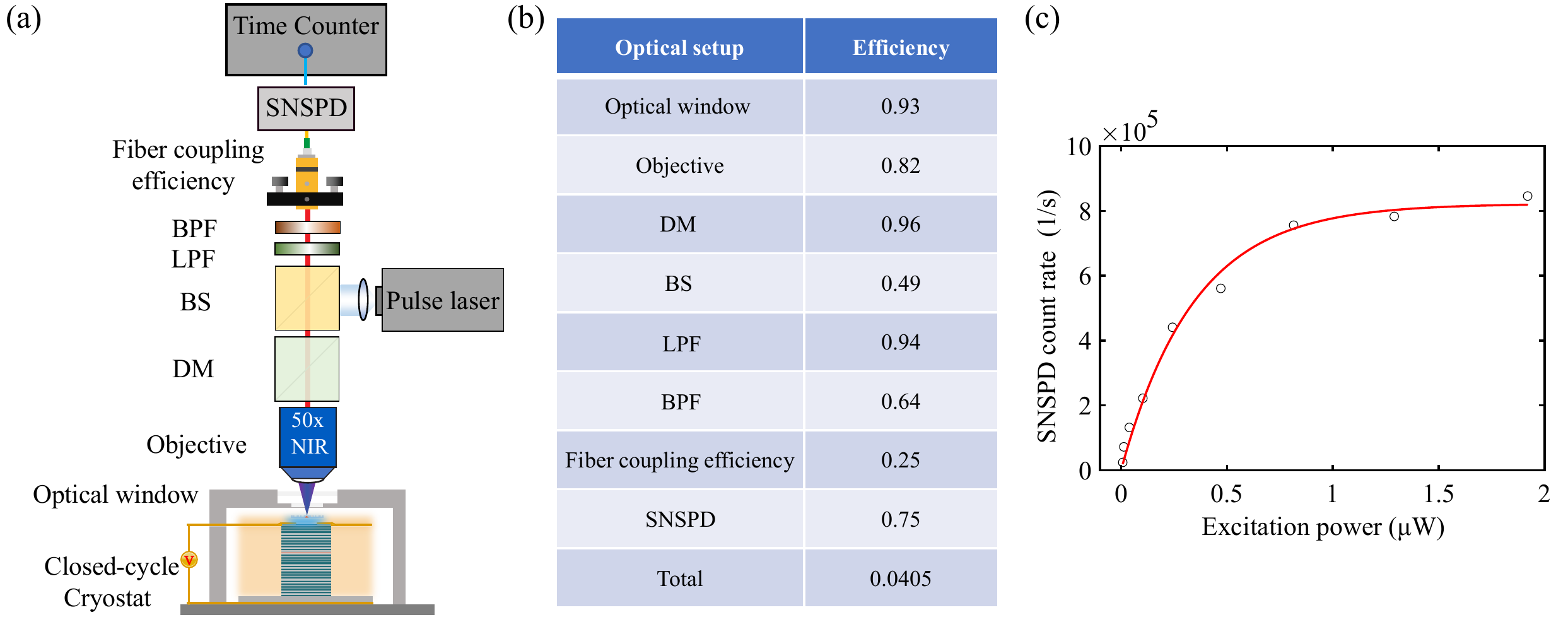}
	\caption{Estimation of the extraction efficiency at the first lens. (a) Setup for the extraction efficiency measurement. (b). Transmissions of the elements in the setup. (c) Photon count rates as a function of the pulsed laser power.}
	\label{SIfig:Fig4}
	%\end{center}
\end{figure}

\newpage
\section{Scheme for on-chip resonant excitation}
Inspired by recent works of exploiting asymmetrical microcavities for resonance fluorescence~\cite{wang2019towards,gerhardt2019polarization}, we propose to employ an elliptical micropillar for obtaining a single-mode laser with H-polarization, as shown in Fig.~S11(a). Meanwhile an asymmetrical CBG is employed in which a charged exciton is resonant with V-polarized cavity mode and the tail of H-polarized cavity mode is used for the resonant excitation, as shown in Fig.~S11(b). In such a configuration, an external polarizer along V direction as shown in Fig.~S11(c) will remove the H-polarized laser background and result in V-polarized single photons with a high indistinguishability.

\begin{figure}[!h]
	%\begin{center}
	\includegraphics[width=\linewidth]{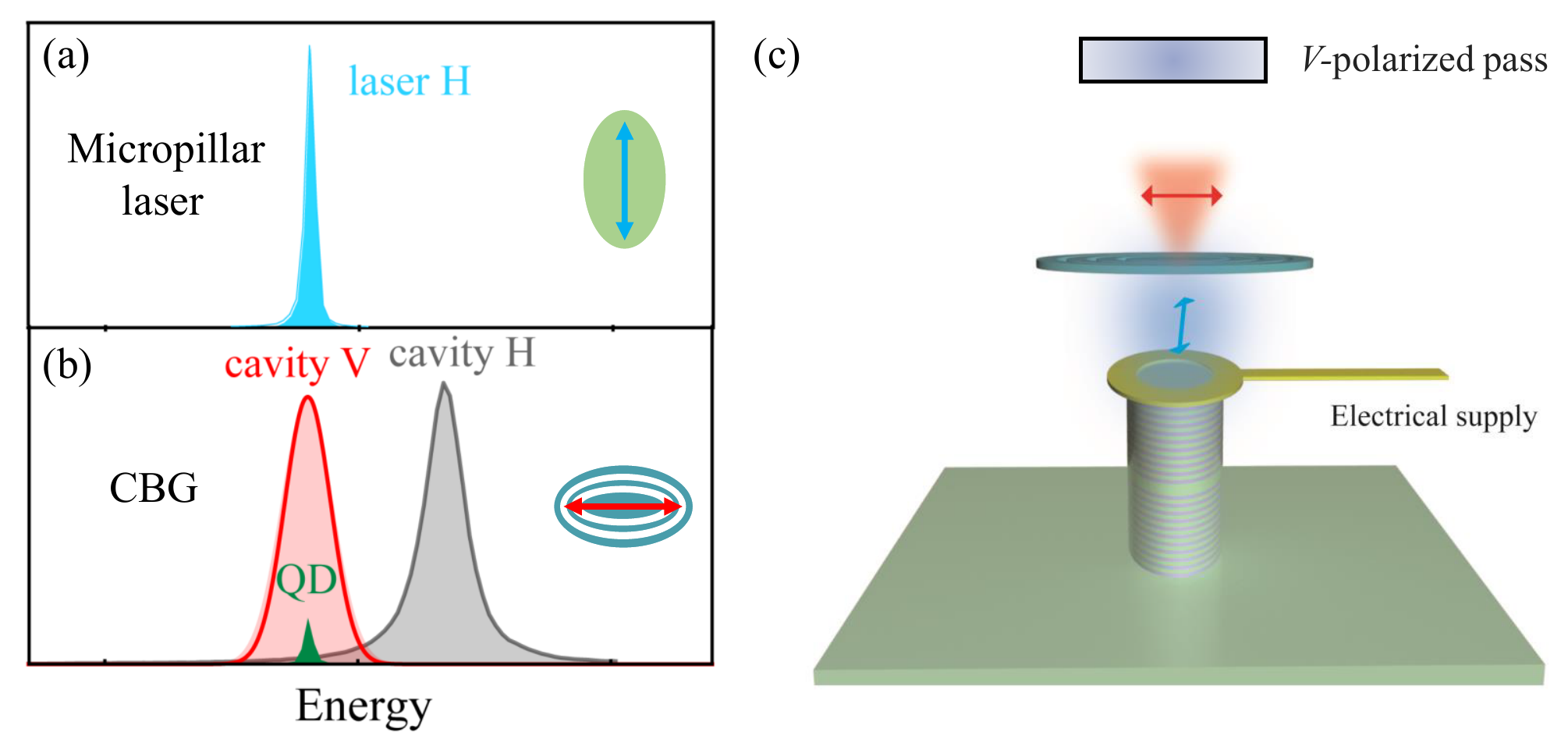}
	\caption{Implementation of on-chip resonant excitations in our platform. (a) H-polarized lasing mode from an elliptical micropillar. (b) Spectral alignment for QD and cavity modes of an asymmetrical CBG for the resonant excitation. (c) Schematic of the on-chip resonant excitation scheme.}
	\label{SIfig:Fig11}
	%\end{center}
\end{figure}

\newpage
\section{Comparison to devices in the literature}
We’ve compared our devices with representative of the state-of-the-art in Table~1. Compared to the state-of-the-art devices under on-chip excitation, our device shows a much better single-photon purity and a higher extraction efficiency. However, the best reported devices under off-chip excitation still exhibit better metrics in terms of extraction efficiency, single-photon purity as well as the indistinguishability. We believe that the performance of our device could be further significantly improved with optimizations of the fabrication process and implementations of on-chip resonant excitation in Fig.~S11.
\begin{figure}[!h]
	%\begin{center}
	\includegraphics[width=0.9\linewidth]{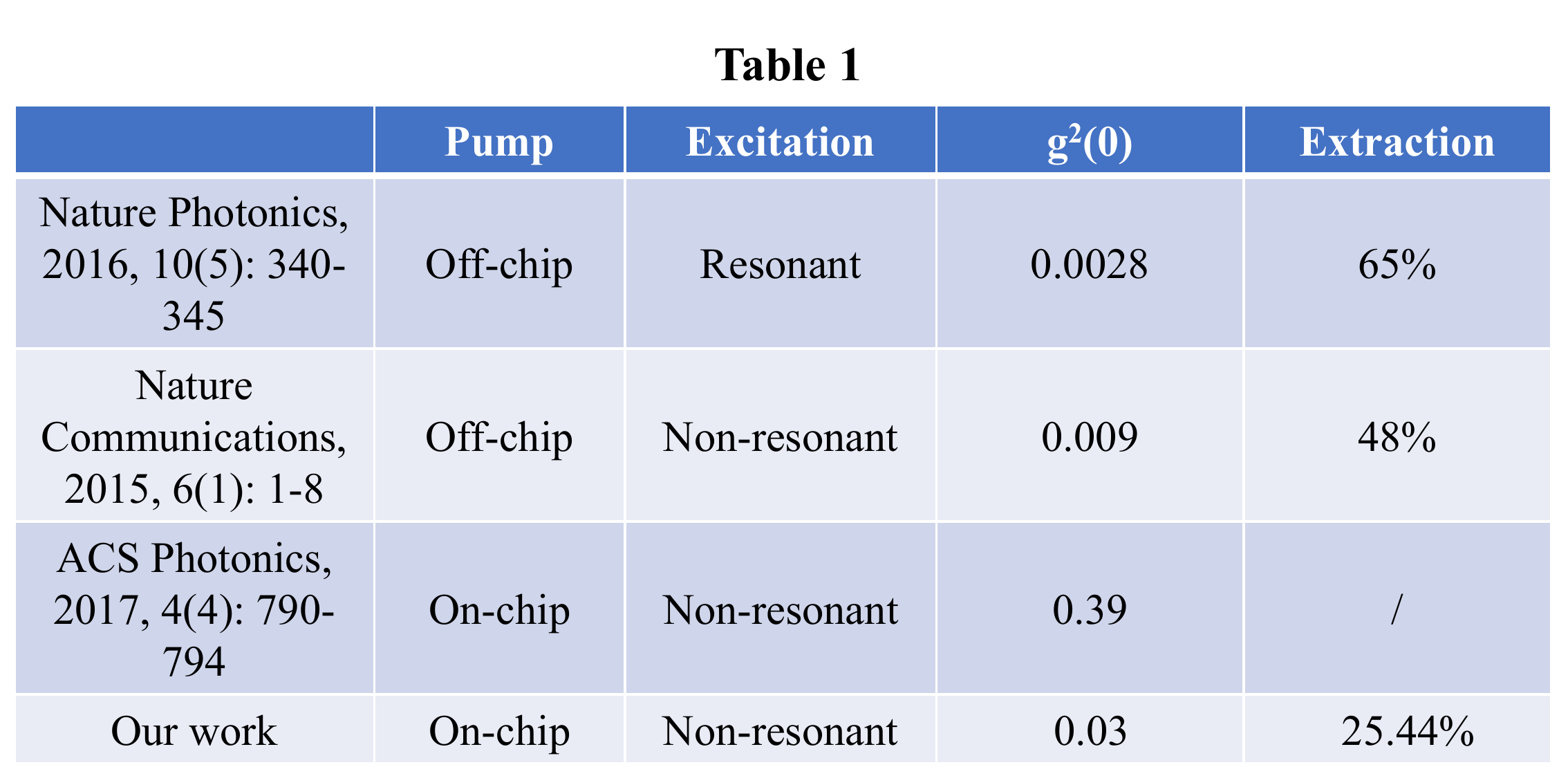}
	\caption{Comparison to the devices in the literature.}
	\label{SIfig:Fig11}
	%\end{center}
\end{figure}

\end{document}